\documentclass[showpacs,prl,twocolumn]{revtex4}
\usepackage{amssymb}
\usepackage{graphicx}
\usepackage{dcolumn}
\usepackage{bm}
\usepackage{amsmath}

\begin{document}

\title{Self-Focusing Dynamics of Coupled Optical Beams}
\author{Amiel A. Ishaaya, Taylor D. Grow, Saikat Ghosh, Luat T. Vuong and Alexander L. Gaeta}
\address{School of Applied and Engineering Physics, Cornell University, Ithaca, NY 14853, USA}

\begin{abstract}
We theoretically and experimentally investigate the mutual collapse
dynamics of two spatially separated optical beams in a Kerr medium.
Depending on the initial power, beam separation, and the relative
phase, we observe repulsion or attraction, which in the latter case
reveals a sharp transition to a single collapsing beam. This
transition to fusion of the beams is accompanied by an increase in
the collapse distance, indicating the effect of the nonlinear
coupling on the individual collapse dynamics. Our results shed light
on the basic nonlinear interaction between self-focused beams and
provide a mechanism to control the collapse dynamics of such beams.
\end{abstract}

\pacs{42.65.Jx, 42.65.Sf}

\maketitle

Nonlinear collapse to a singularity and the accompanying richness of
physical effects is observed in several fields of physics, such as
optics \cite{Berge0}, hydrodynamics \cite{Zeff}, and plasmas
\cite{Robinson}. Self focusing and collapse of optical beams with a
power above a critical power $P_{cr}$ \cite{Fibich} in nonlinear
Kerr media was first predicted in the 1960's \cite{Kelley}. Over the
past decade, with the advent of high peak power ultra-short pulsed
lasers, self-focusing dynamics of optical beams have revealed a
remarkable richness of spatial and temporal nonlinear phenomena.
These include observation of the universal self-similar spatial
collapse profile known as the {\it Townes} profile \cite{Moll},
multiple filamentation \cite{Braun}, self steepening and pulse
splitting \cite{Fibich1,Gaeta}, multiphoton ionization, and
supercontinuum generation \cite{Gaeta,Alfano}, along with saturation
and plasma generation that typically arrest the collapse
\cite{Mlejnek1}.

Related optical beam interactions within the context of spatial
solitons \cite{Stegeman, Meier} have drawn considerable interest in
recent years. Spatial solitons in 1D Kerr media are shown to
interact in a particle-like elastic manner, where the number of
solitons and the corresponding directions and propagation velocities
are conserved \cite{Stegeman, Zabusky}. Depending on the relative
phase, attractive and repulsive forces and power transfer are
observed between interacting solitons \cite{Gordon}. In saturable
nonlinear media, which can support (2+1)D solitons, phenomena such
as soliton fusion, fission, annihilation, and spiraling occur
\cite{Gatz,Tikhonenko,Krolikowski,Mitchell}. The interaction between
incoherent solitons, where the medium responds non-instantaneously,
exhibits similar effects as those observed with coherent solitons in
nonlinear saturable media \cite{Andersen}.

In the regime of optical beam collapse in a Kerr medium, where self
focusing is dominant and beams do not maintain their spatial
distribution, only a few theoretical studies of beam interactions
have been reported \cite{Landman,McKinstrie,Berge,Berge1}, and to
the best of our knowledge no experiments have been performed. While
in these initial theoretical studies several qualitative trends,
such as repulsion, attraction, and fusion of two beams, were
identified, the detailed dynamics and the transition to fusion of
two beams, especially when each beam has a power near $P_{cr}$, has
not been explored.

In this Letter, we investigate both theoretically and experimentally
the spatial collapse dynamics of two coherently coupled beams in
Kerr media. We observe repulsion and attraction between the
collapsing beams, and a sharp transition to fusion of the beams,
which is dependent on their initial power, spatial separation and
relative phase. We further show that this transition, accompanied by
a peak in the collapse distance, can be exploited to control and
manipulate the collapse dynamics of two coupled beams. Our results
shed light on the basic nonlinear interaction between self-focused
collapsing beams and are applicable in different scenarios,
including that of multiple filamentation and collapse of complex
multi-lobe beams such as necklace beams \cite{Soljacic}.

To investigate the propagation and collapse of two spatially
separated beams in Kerr media, we numerically integrate the scalar
(2+1)D nonlinear Schr\"{o}dinger equation (NLSE), neglecting
dispersion and high-order terms. These assumptions are reasonable as
long as the dispersion length is longer than the nonlinear and
diffraction length scales. We express the (2+1)D NLSE in normalized
units as

\begin{equation}
i\psi_{\zeta}+\frac{1}{4}\nabla^{2}_{\perp}\psi+\frac{L_{df}}{L_{nl}}|\psi|^2\psi=0,\label{NLSEnon}
\end{equation}
where $\psi$ is proportional to the amplitude
$A(x,y,z)=A_{0}\psi(\mu,\nu,\zeta)$ of the envelope of the electric
field, $\zeta=z/L_{df}$ is the normalized coordinate in the
propagation direction, $\mu=x/r_0$ and $\nu=y/r_0$ are the
normalized transverse coordinates, $r_0$ is the characteristic
radius of the input beam,
$\nabla^{2}_{\perp}=\partial^2/\partial\mu^2+\partial^2/\partial\nu^2$
is the transverse Laplacian, $L_{df}=kr_0^2/2$ is the diffraction
length, $L_{nl}=n_0n_2c|A_{0}|^2/\lambda$ is the nonlinear length,
$k = 2\pi n_0/\lambda$ is the wave number, $\lambda$ is the vacuum
wavelength, $n_0$ is the linear index of refraction, and $n_2$ is
the nonlinear index coefficient. When the total power
$P=(n_0c/2\pi)|A_{0}|^2\int|\psi|^2d\mu d\nu$ in the beam is greater
than $P_{cr}=\alpha(\lambda^2/4\pi n_0n_2)$, where $\alpha$ is a
constant that depends on the initial spatial profile \cite{Fibich},
the self-focusing term dominates and the beam undergoes collapse.

In the present study, we assume an initial field composed of two
spatially separated, parallel, Gaussian beams of the form
$\psi(\mu,\nu,0)=C\{e^{-[(\mu-\Delta_0/2)^2+
\nu^2]}+e^{i\varphi}e^{-[(\mu+\Delta_0/2)^2+\nu^2]}\}$, where
$\Delta_0$ is the normalized initial spatial separation of the
beams, $\varphi$ is the relative phase between the beams, and $C$ is
a normalization constant that depends on the total power. To exclude
cases where the two beams are initially unresolvable and hence
collapse as a single beam, we assume $\Delta_0>1$. Since a
collapsing Gaussian beam is known to form a {\it Townes}
distribution carrying one critical power \cite{Moll}, we limit our
investigation to the regime of powers $2P_{cr}<P<10P_{cr}$. This
corresponds to powers $P_{cr}<P_G<5P_{cr}$ for each beam, thus
enabling their individual collapse, but restricts the collapse
distance to be larger than $0.25L_{df}$.

For a fixed total power of $2.5P_{cr}$, various initial separations
$\Delta_0$, and for a relative phase of either 0 or $\pi$ (in- and
out-of-phase cases), we integrate Eq. (\ref{NLSEnon}) with the
split-step method and obtain the intensity distribution as a
function of $\zeta$. Just before collapse, when the peak intensity
reaches 50 times the initial peak intensity, we record the $\zeta$
value and the separation between the two collapsing peaks. The
distance between the collapse points $\Delta_{col}$ and the collapse
distance $\zeta_{col}$ as functions of the initial separation are
shown in Figs. 1(a) and 1(b), respectively. When the separation is
large, for both the in- and out-of phase cases, the collapse
dynamics of each beam are essentially independent of the other beam,
and the distance between the collapse points is equal to the initial
separation. For the in-phase case, as the initial separation is
decreased the collapsing beams are attracted towards each other, and
their separation decreases with respect to the initial separation.
At some critical separation ($\Delta_0\approx1.9$) a sharp
transition to one collapsing beam is observed. This transition to
fusion of the two beams is accompanied by more than a $40\%$
increase in the collapse distance with respect to independent beam
collapse. The lengthening of collapse distance represents a critical
slowing down, associated with the phase transition between
independent collapse and fusion. For the out-of-phase case the
behavior is drastically different. When the initial separation is
decreased the collapsing beams repel each other and do not undergo
fusion to a single collapsing point. This resembles the well-known
repulsion observed between two anti-phased spatial solitons
\cite{Stegeman}. Here the collapse distance is decreased with regard
to independent beam collapse.

\begin{figure}[t]
\centering
\includegraphics[width=6cm]{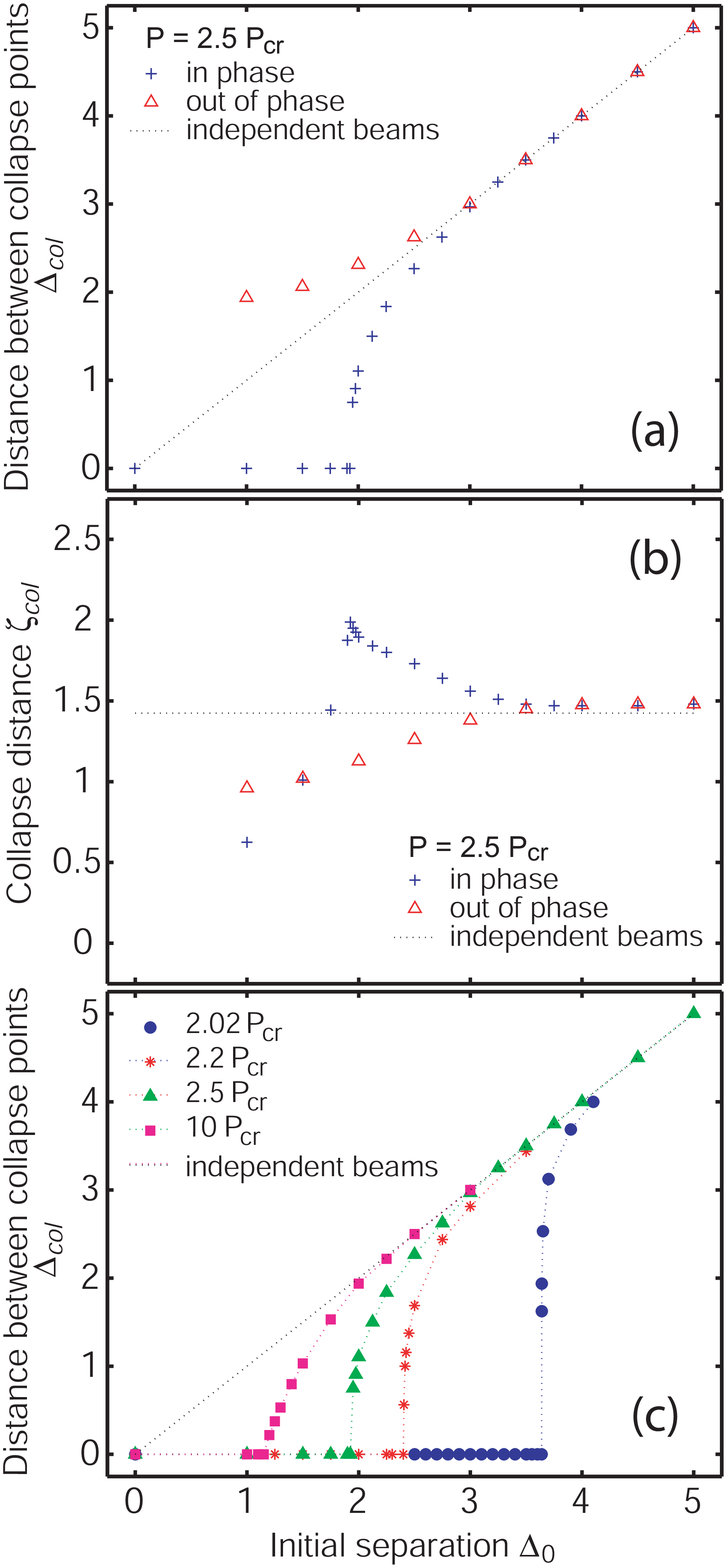}
\caption{(Color online) Numerical calculations of the collapse
dynamics of two coupled Gaussian beams in Kerr media. (a) Spatial
separation between collapse points and (b) collapse distance as a
function of initial separation between the beams, for 0 and $\pi$
relative phase (total power $2.5P_{cr}$); (c) same as the in-phase
case in (a) for various input powers.}
\end{figure}

We further investigate for the in-phase case the dependence of the
above transition on the total power. Figure 1(c) shows the distance
between collapse points as a function of initial separation for
different total powers. As the total power is decreased, approaching
$2P_{cr}$ (i.e. one critical power per beam), the critical
separation at which the transition occurs increases, and the
transition becomes steeper. With a total power of $2.02P_{cr}$, the
critical separation is almost three times that for a total power of
$10P_{cr}$. As $P\rightarrow P_{cr}$, the distance to collapse is
extended, allowing for longer lengths of mutual attraction and
eventually for fusion into one beam (e.g. with a total power of
$2.5P_{cr}$ the independent collapse distance is $1.4L_{df}$ while
with $2.02P_{cr}$ it is $7L_{df}$, and is more than doubled at the
transition region to $18L_{df}$). At relatively large total powers,
the transition occurs at smaller initial separations, such that the
beams appear to be decoupled for most initial separations. Unlike
the case of soliton interactions, these dynamics are governed
simultaneously by two competing processes: individual beam collapse
and fusion of the beams.

To experimentally investigate the collapse dynamics of this system,
we use the arrangement schematically shown in Fig. 2. A pulsed,
Ti:Sapphire amplified laser beam, with 1-kHz, 90-fs pulses centered
at 800 nm, is passed through a spatial filter to obtain a nearly
Gaussian profile. This beam is sent into a two-arm interferometer,
composed of a 50:50 beam splitter and two retro reflectors, such
that the two spatially separated beams can be recombined at the same
beam splitter. The separation of the two beams is changed by moving
one of the retro reflectors in the transverse direction. The
temporal alignment needed for coherently combining two short pulses
and for controlling their relative phase is achieved by changing the
length of one arm with a high-precision motorized stage. The
combined beams emerging from the interferometric setup are
telescoped down to obtain a 0.29-mm beam waist ($\omega_{0}$) and
then sent into a 28-cm rod of BK7 glass (equivalent to
$0.56L_{df}$). The intensity distribution at the input and output
faces of the BK7 rod are imaged onto a 12-bit CCD camera. In order
to monitor the relative phase of the two combined beams, the
far-field intensity fringes of the input field distribution are
imaged onto an additional CCD camera.

\begin{figure}[b]
\centering\includegraphics[width=7.5cm]{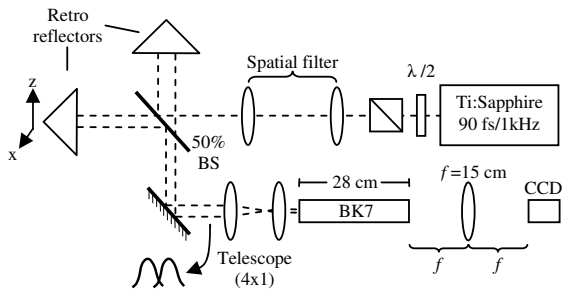}\caption{Ultrafast
experimental arrangement for self focusing and collapse of two
spatially separated Gaussian beams in BK7 glass.}
\end{figure}

Initially, one arm is blocked, and the laser power is adjusted so
that collapse of a single beam is obtained at the output face of the
BK7 rod. The energy per pulse is measured to be 3.6 $\mu$J,
equivalent to 40 MW, which is about a factor of 10 higher than
expected from the CW numerical calculations for collapse after
$0.56L_{df}$ in BK7 glass. We attribute this discrepancy in powers
mainly to dispersive effects in the long glass rod and find that
this translates into merely a scaling factor for the power and does
not affect significantly the spatial dynamics under investigation.
This power, at which collapse of a single beam is observed at the
output, is held constant throughout the experiments. In practice,
due to coherent overlap of the beams on the beam splitter, the total
power slightly increased as their separation is decreased. For beam
separations greater than $1.5\omega_{0}$ this increase is calculated
to be less than $32\%$.

\begin{figure}[b]
\centering\includegraphics[width=7cm]{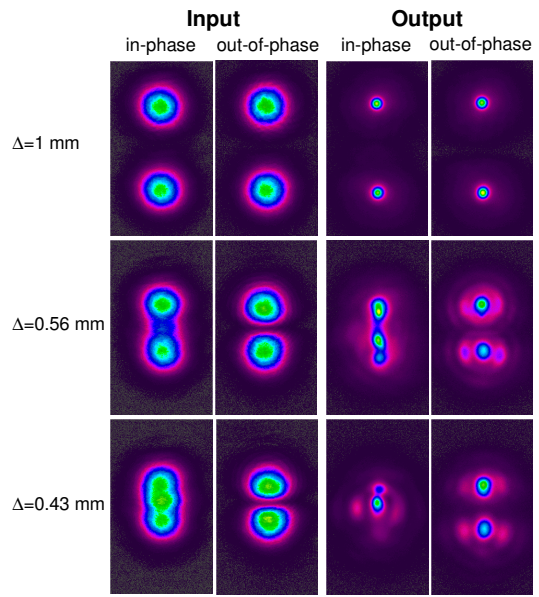}\caption{(Color
online) Experimental input and output intensity distributions of two
collapsing beams with different initial separations for the in- and
out-of-phase cases. The Gaussian beam waist is 0.29 mm, the total
energy for the two beams is 7.2 $\mu$J, and the propagation distance
in the BK7 glass sample is 28 cm.}
\end{figure}

After setting the power, we unblock the two arms and for each
initial separation of the two beams we record the input and output
intensity distributions. This is done for the case in which the
beams are temporally misaligned (i.e. the independent beam case) and
for both the in- and out-of-phase cases, where the relative phase is
zero and $\pi$, respectively. Several representative intensity
distributions for different initial separations and relative phase
are shown in Fig. 3. For the in-phase case, the observed behavior is
in agreement with our simulations in which two collapsing points
occur at the output when the initial separation is large, whereas
only one collapse point is seen when the initial separation is
small. This is a clear indication of predicted attraction and fusion
between two collapsing in-phase beams. For the out-of-phase case,
two collapsing points are always observed at the output, indicating
repulsion between the collapsing beams, which is also in agreement
with our theoretical results.

\begin{figure}[t]
\centering\includegraphics[width=6cm]{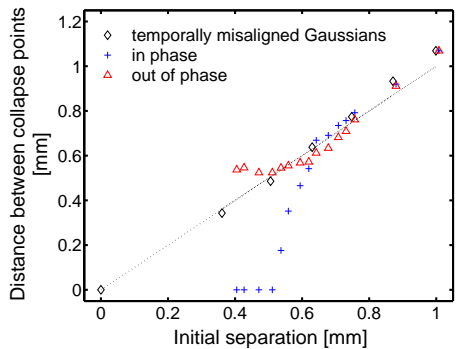}\caption{(Color
online) Experimental measurements of the spatial separation between
collapse points as a function of initial spatial separation between
two beams for 0 and $\pi$ relative phase and for the temporally
misaligned case. Here the initial beam waist is 0.29 mm, the total
energy in the two beams is 7.2 $\mu$J, and the collapsing beams are
measured after 28 cm of BK7 glass, which corresponds to the distance
where independent collapse occurs.}
\end{figure}

From the recorded intensity distributions, we measure the initial
separation between the beams at the input face and the corresponding
separation between the collapse points at the output face, and the
results are plotted in Fig. 4. As expected from the numerical
calculations, attraction between the beams and a sharp transition to
a single collapse point is observed for the in-phase case. This
occurs at a separation distance of $1.75\omega_{0}$, corresponding
to the numerical calculations with $P_{total}\approx4P_{cr}$ [see
Fig. 1(c)], which is in good agreement with the collapse distance.
For the out-of-phase case, the beams repel each other, and the
separation between the collapse points does not decrease below 0.52
mm, which is equal to $1.8\omega_{0}$.

In order to experimentally test our ability to dynamically control
the collapse of two coupled beams via their relative phase, we
slowly sweep the time overlap of the two beams (near the optimal
overlap point), thus changing their relative phase periodically.
This is performed for two different initial separations, the first
at a separation of 0.41 mm corresponding to $1.4\omega_{0}$ where
the in-phase case results in fusion into one collapse point, and the
second at a separation of 0.54 mm corresponding to $1.86\omega_{0}$
where the distance between the two collapse points is significantly
affected by the relative phase. In the first case the output
distribution oscillates between two distinct states with either one
or two collapse points. In the second case, the output distribution
always has two collapse points, for which the separation increases
and decreases periodically. When the relative phase is in the
intermediate state (neither 0 or $\pi$), the power appears to
redistribute between the collapse points in an analogous way to that
observed in 1D soliton experiments \cite{Meier}.

In conclusion, we have investigated the collapse dynamics of two
spatially separated coupled beams in a Kerr medium. For the in-phase
case, we experimentally observe a sharp transition to fusion of the
two collapsing beams, which depends on the initial separation and
power. We predict that this transition is accompanied by a longer
collapse distance with respect to independent collapse. For the
out-of-phase case, the beams repel each other and collapse
independently. Our results of two-beam interaction serve as a basis
for analyzing more complicated scenarios where multiple beams or
composite multi-lobe beams undergo self focusing and collapse.

We acknowledge unpublished related work of K. D. Moll, funding from
the National Science Foundation under grant No. PHY-0244995, and
from the Army Research Office under grant No. 48223-PH.

\end{document}